\documentclass[pra,twocolumn, preprintnumbers,amsmath,amssymb, nofootinbib, superscriptaddress]{revtex4}

\usepackage{amsmath}
\usepackage{amssymb}
\usepackage{amsthm}
\usepackage{graphicx}
\usepackage{pstricks}
\usepackage{bm}
\usepackage{hyperref}

\newcommand{\colo}{\,\hbox{:}\,}              
\newcommand{\ad}{^\dagger }           
\newcommand{\avg}[1]{\langle #1\rangle }  
\newcommand{\bld}[1]{\boldsymbol{#1}}
\newcommand{\bra}[1]{\langle #1|}     
\newcommand{\dya}[1]{\ket{#1}\bra{#1}}           
\newcommand{\id}{I}			          
\newcommand{\ii}{\mathrm{i}}			  
\newcommand{\ket}[1]{|#1\rangle}     
\newcommand{\od}{\odot }             
\newcommand{\omt}{\omega }          
\newcommand{\ot}{\otimes }           
\newcommand{\pp}{p}                  
\newcommand{\ra}{\rightarrow }       
\newcommand{\Tr}{{\rm Tr}}                        
\newcommand{\vb}{\,|\,}              
\newcommand{\vect}[1]{\bm{#1}}       

\newcommand{\Dfunc}{\mathbb{D}}    
\newcommand{\Dmat}{\mathbf{D}}     

\newcommand{\Omat}{\mathbf{O}}

\newcommand{\Rmat}{\mathbf{R}}

\newcommand{\Smat}{\mathbf{S}}
\newcommand{\Sbmat}{\mathbf{\bar S}}

\newcommand{\ta}{t}
\newcommand{\Tmat}{\mathbf{T}}
\newcommand{\Tsup}{\mathcal{T}}

\newcommand{\CQT}{RBGriffiths:ConsistentQuantumTheory}

\long\def\ca#1\cb{}

 \def\outl#1{}  \def\xa{} \def\xb{}  

\ca
 \def\outl#1{\par{\medskip\noindent\hspace*{.5cm}\bf
      \mathversion{bold}#1\mathversion{normal}\smallskip} }
 \long\def\xa#1\xb{}
 
\cb

 \def\outl#1{\par{\medskip\noindent\hspace*{.5cm}\bf
      \mathversion{bold}#1\mathversion{normal}\smallskip} }
 \def\xa{} \def\xb{}  

\newcommand{\al}{\alpha }
\newcommand{\bt}{\beta }
\newcommand{\gm}{\gamma }

\newcommand{\dl}{\delta }
\newcommand{\Dl}{\Delta }

\newcommand{\zt}{\zeta }
\renewcommand{\th}{\theta } 

\newcommand{\kp}{\kappa }
\newcommand{\lm}{\lambda }

\newcommand{\sg}{\sigma }

\newcommand{\om}{\omega }

\newcommand{\HC}{\mathcal{H}}

\newcommand{\TC}{\mathcal{T}}

\begin{document}

\title{Consistent histories for tunneling molecules subject to collisional decoherence}
\author{Patrick J. Coles}
\affiliation{Department of Physics, Carnegie Mellon University, Pittsburgh, Pennsylvania 15213, USA}
\affiliation{Centre for Quantum Technologies, National University of Singapore, Singapore}
\author{Vlad Gheorghiu}
\affiliation{Department of Physics, Carnegie Mellon University, Pittsburgh, Pennsylvania 15213, USA}
\affiliation{Institute for Quantum Information Science and Department of Mathematics and Statistics, University of Calgary, Calgary, Alberta T2N 1N4, Canada}
\author{Robert B. Griffiths}
\affiliation{Department of Physics, Carnegie Mellon University, Pittsburgh, Pennsylvania 15213, USA}

\xa
\begin{abstract} 
  The decoherence of a two-state tunneling molecule, such as a chiral molecule
  or ammonia, due to collisions with a buffer gas is analyzed in terms of a
  succession of quantum states of the molecule satisfying the conditions for a
  consistent family of histories. With $\hbar\omega$ the separation in energy
  of the levels in the isolated molecule and $\gamma$ a decoherence rate
  proportional to the rate of collisions, we find for $\gamma \gg \omega$
  (strong decoherence) a consistent family in which the molecule flips
  randomly back and forth between the left- and right-handed chiral states in
  a stationary Markov process.  For $\gamma < \omega$ there is a family in
  which the molecule oscillates continuously between the different chiral
  states, but with occasional random changes of phase, at a frequency that
  goes to zero at a phase transition $\gamma = \omega $.  This transition is similar to the
  behavior of the inversion frequency of ammonia with increasing pressure,
  but will be difficult to observe in chiral molecules such as D$_2$S$_2$.
  There are additional consistent families both for $\gamma > \omega $ and for $\gamma <
  \omega $.  In addition we relate the speed with which chiral information is
  transferred to the environment to the rate of decrease of complementary types
  of information (e.g., parity information) remaining in the molecule itself.
\end{abstract}
\xb 

\maketitle

\section{Introduction\label{sct1}}
\xa

The decoherence produced by the interaction of a quantum system with its
environment is ubiquitous in nature and plays an important role in current
quantum theory in at least two ways.  First, it is widely believed that
decoherence helps understand how the classical physics of macroscopic objects
emerges as an approximation to underlying quantum mechanical laws.  Second,
decoherence is the great enemy of quantum computation, quantum cryptography,
and other schemes seeking to utilize specifically quantum effects for
particular processes.  For both reasons it is important to study specific
microscopic models from which one can hope to obtain general principles for
decoherence.  The present paper is the study of a simple two-level system
which can be thought of as a crude microscopic model of chiral molecules or
ammonia in which the lowest quantum energy levels correspond to the nearly degenerate eigenstates of a double-well potential, with decoherence occuring through collisions with particles in the environment.

Microscopic studies of decoherence are often framed in terms of a master
equation for the density operator of the decohering system.  Such descriptions
are perfectly valid, but because they represent the average of a large
ensemble of nominally identical systems, each with a different specific time
development, they provide less information and less physical insight than the
actual history of a single system.  For example, in the phenomenon of
intermittent fluorescence a single ion in a trap shows intermittent light and
dark periods when it does or does not scatter resonance radiation
\cite{PlKn98}.  This behavior is not directly
reflected in the density operator, even though from the latter one can deduce
parameters which govern the statistical behavior of the individual ion.

Another way to understand the limitations of the density operator description
is to consider its classical analog for a Brownian particle confined to a
small but macroscopic volume of a fluid by rigid walls.  The probability
distribution $\rho(\vect{r},t)$ of the particle position $\vect{r}$ will
eventually tend to a constant over the region accessible to the particle,
whereas the particle itself will continue to exhibit a sort of random walk.
More details of what is going on in this steady-state situation is provided by
the joint probability distribution of the sequence of successive positions
$\vect{r}_1,\,\vect{r}_2,\ldots$ of the particle at a sequence of times
$t_1,\,t_2,\ldots$, that is, its \emph{history}. Averaging over a large number
of histories will yield $\rho(\vect{r},t)$, but in the process the information
needed for a more detailed temporal description of the particle is lost. In the quantum case unravelings of the master equation provide a more
detailed description of the microscopic time development, but these are often
viewed as mathematical artifacts having no necessary
connection with what is really going on in the quantum system.  There are many
possible unravelings; which, if any are correct?  Standing in the way of
answering this question is the infamous \emph{measurement problem} of quantum
foundations: textbook quantum mechanics introduces probabilities by means of
measurements, but cannot say what it actually is that is being measured.  

However, the \emph{consistent histories} or \emph{decoherent
  histories}---hereafter simply referred to as \emph{histories}---formulation
of quantum mechanics, has no measurement problem, and provides the tools
needed to identify trajectories or sequences of events that actually correspond to physical
processes.  Or, putting it another way, it allows one to identify certain
classes of microscopic stochastic processes which can be consistently
described in a fully quantum mechanical terms. The histories approach
has previously been applied to quantum optical systems by Brun \cite{Brn02, PhysRevLett.78.1833}, though we believe the material
presented here is the first application to the case of tunneling molecules,
including chiral molecules.

Early in the development of quantum mechanics the question arose as to why
chiral molecules are observed in left- and right-handed versions even though
the quantum ground state should be a symmetrical combination of the two forms.
Hund \cite{Hund1927} provided the first step in addressing this paradox when
he pointed out that the two enantiomers correspond to the two wells of a
symmetrical potential with two minima, and that the time required to tunnel
from one well to another for a typical chiral molecule is extremely long.  A
second step was provided by Simonius \cite{PhysRevLett.40.980} who observed
that interaction with the environment of a suitable sort (i.e., decoherence,
though when he wrote that term was not yet current) can stabilize the chiral
states for periods substantially longer than the tunneling time.  At present
it seems widely accepted that such decoherence is an important aspect of the
stability of chiral molecules, though there have been dissenting voices, e.g.
\cite{PhysLettA.147.411}.

The time dependence of the two-state model introduced in Sec.~\ref{sct2}, when
analyzed in terms of consistent histories using the principles discussed in
Sec.~\ref{sct3}, and applied to specific consistent families in
Sec.~\ref{sct4}, yields some insight into this stability problem.  In
particular, we find that if the rate of decoherence due to collisions $\gm$ (a
parameter in our model) is much larger than the tunneling rate $\om$ in an
isolated molecule, there is a consistent family in which the molecule spends a
long but random period of time in each of the chiral states before flipping to
the one of opposite chirality, in a two-state Markov process.  As $\gm$
decreases the flips become more rapid and the ``dressed'' quantum states
between which the flips occur become less and less chiral, with this type of
family finally disappearing at a phase transition $\gm = \om$. For $\gm < \om$
there is a different consistent family with a rapid but continuous oscillation
of the molecule back and forth between its chiral states, interrupted at
random times by a change in phase.  There are a variety of other consistent
families, and these are discussed, along with their physical interpretation,
in Sec.~\ref{sct4}.  Most chiral molecules in most circumstances will be in
the strong decoherence regime. We give some approximate numerical values in
Sec.~\ref{sbct6.1} for D$_2$S$_2$ in a buffer gas of helium, as it has been
the subject of some careful calculations in \cite{PhysRevLett.103.023202}.  On
the other hand the ammonia molecule, which though not itself a chiral molecule
can behave like one in certain rotational states, has an inversion (tunneling)
transition with a frequency that goes to zero with increasing pressure. This
is probably an example of, or at least very similar to, the $\gm = \om$ phase
transition, for reasons discussed in Sec.~\ref{sbct6.2}.

An intuitive way of characterizing the stochastic time development in the histories formalism is to quantify the information dynamics, e.g., the loss over time of information about the system's initial state due to random hopping. Such information loss is ultimately connected to decoherence \cite{Jsao03}, which can alternatively be viewed as flow of information about the system to the environment \cite{Zrk03}, and in Sec.~\ref{sct5} we illustrate the quantitative connection between these two views of decoherence for our model. In our model decoherence corresponds to a flow of chiral information---i.e.,
is the molecule left or right handed?---to the environment. In Sec.~\ref{sct5}
we analyze this using quantitative measures defined in Sec.~\ref{sbct3.3}, and
compare the flow of chiral information to the environment with the decrease of
complementary types of information (e.g., parity information) about the earlier state of the molecule
that remain in the molecule itself at later times.

Our conclusions are summarized in Sec.~\ref{sct7}, which also indicates some
ways in which the results reported here could be usefully extended.  A few
mathematical derivations and details are placed in appendices. 

\xb
\section{Microscopic model and master equation\label{sct2}}
\xa

\subsection{Double-well potential and collisions\label{sbct2.1}}

We consider a quantum system, the molecule, with a double-well potential in
which the two lowest energy eigenstates, $\ket{0}$ (even parity) and $\ket{1}$
(odd parity), are sufficiently well separated in energy from all the higher
levels that the latter can be ignored.  The Hamiltonian is of the form
\begin{equation}
  H = (1/2)\hbar\omt Z,
\label{eqn1}
\end{equation}
where $Z = \dya{0} - \dya{1}$ is the Pauli operator $\sg_3=\sg_z$, so the
energy splitting between the levels is $\hbar\omt$.  The linear combinations
\begin{equation}
\ket{R}=\bigl(\ket{0}+\ket{1}\bigr)/ \sqrt{2},\quad 
\ket{L}=\bigl(\ket{0}-\ket{1}\bigr)/ \sqrt{2},\quad 
\label{eqn2}
\end{equation}
represent the right- and left-handed chiral forms of the molecule, or in
ammonia the nitrogen on one or the other side of the plane formed by the
hydrogens. In real molecules there are, of course, additional degrees of
freedom---rotations, vibrations, etc.  We are assuming that for our purposes
these can be ignored, i.e., the Hilbert space can be approximated as a tensor
product of these other degrees of freedom with the two levels representing the
tunneling, with negligible coupling between them.  Hence the isolated molecule
can be thought of as oscillating or tunneling between the $\ket{R}$ and
$\ket{L}$ states at an angular frequency $\omt$.  In the Bloch sphere picture
the kets $\ket{R}$ and $\ket{L}$ correspond to the points on the positive and
negative $x$ axis, and the sphere rotates about the $z$ axis as time
increases.

Next we assume the molecule collides randomly with other particles (atoms or
molecules), and the duration of each collision is short compared to the times
we are interested in.  Successive collisions need not be independent of each
other, but we assume that correlations die away rapidly after some
\emph{correlation time} $\tau_c$, which could be shorter than the average time
between collisions in a dilute gas, but might be significantly longer in a
dense gas or liquid.  We will consider properties of the molecule at a
succession of times $t_0$, $t_1$, $t_2$\dots, where the $m$'th time interval,
$\Dl t_m=t_{m+1}-t_m$ is always greater than $\tau_c$, and ideally should be
significantly greater than $\tau_c$. That is, we are using a description which
is coarse-grained in time; the importance of this will appear later.  During
the $m$'th time interval there may be zero or one or more collisions of other
particles with the molecule, and different collisions can have different
effects.  We shall assume that the probability distribution for these
collisions in a particular interval, both for the times at which they occur
and the effects which they have on the molecule, are \emph{statistically
  independent} of what happens in other intervals. Obviously this cannot be
exactly correct, but on physical grounds it seems reasonable provided $\Dl t_m$
is not too short, which is why we assume that it is larger than $\tau_c$. In
addition we assume, as is appropriate for a steady state situation, that this
probability distribution depends only on the length $\Dl t_m$ of the interval,
and not otherwise on $t_m$.

The next assumption is that at the beginning of a time interval of length
$\Dl t$ the molecule and the environment can be adequately described, for the
purposes of what happens next, as a tensor product of a molecule state and
some density operator for the environment.\footnote{Working out the general connection of decoherence with thermodynamic irreversibility and the
  properties of steady state is, at a fundamental level, an unsolved problem.
  Our hope is that its eventual resolution will justify present practice by
  the experts in both the quantum information and decoherence communities,
  whose example we are following here. We note that some detailed quantum mechanical treatments of decoherence for situations similar to ours can be found in, e.g., \cite{HornbergerEPL2007, PhysRevB.82.245417} and references therein.}
The latter is the quantum analog of a probability distribution for incoming particles which might collide with the
molecule during this time interval.  This density operator can be ``purified''
by regarding it as arising from an entangled pure state between the
environment and an auxiliary reference system, which we also take to be part
of the environment. The overall time development of the molecule and the environment during the
interval $\Dl t$ is then given by a unitary time development operator,
corresponding to an appropriate Hamiltonian, acting on the system and
environment, resulting in an isometry mapping the Hilbert space $\HC_M$ of the
molecule onto $\HC_M\ot\HC_E$, where $\HC_E$ is the Hilbert space of the
environment.  If one traces out the environment the result is a quantum
operation or channel from $\HC_M$ to itself: the channel input is the molecule
at the beginning of the time interval $\Dl t$ and its output is the molecule at
the end of this interval.  It is represented by a completely-positive
trace-preserving (CPTP) superoperator $\Tsup(\Dl t)$ from the space $\hat\HC_M$
of linear operators on $\HC_M$ to itself. Tracing out the molecule instead of
environment at the end of the time interval results in a corresponding CPTP
map $\TC^c(\Dl t)$ from $\hat\HC_M$ to $\hat\HC_E$, representing the
\emph{complementary} channel.  See, for example, \cite{PhysRevA.83.062338} for further
details on how the direct and complementary channel are related to the
isometry.

The assumption of statistical independence of successive time intervals, and
that the environment is in a steady state, allows us to treat the interval
from $t_1$ to $t_2$ in the same way as the interval from $t_0$ to $t_1$. Thus
a succession of time intervals can be thought of, so far as the molecule is
concerned, as a set of channels in series, with dynamics corresponding to an
appropriate composition of the superoperators $\Tsup(\Dl t_m)$ for the
corresponding intervals.

The use of a superoperator $\Tsup(\Dl t)$ that depends only on the length
$\Dl t$ of the time interval may give rise to the misleading impression that we
are assuming exactly the same number and type of collisions for any interval
of length $\Dl t$. But this is not so.  To understand why, consider a classical
stochastic process for which the independence of successive time intervals
justifies using a Markov model, and for simplicity assume that all the time
intervals are of equal length.  Let $j$ be a discrete index labeling molecule
states at a single time, $j_m$ its value at the beginning of the $m$'th time
interval, and $M^{(n)}(j',j)$ the Markov matrix for a transition $j\ra j'$ if
precisely $n$ collisions occur in one time interval. The probability
distribution for a collection of histories that all begin in the state $j_1$
conditioned on a specified set $n_1,n_2,\ldots n_f$ of numbers of collisions
in the different intervals is
\begin{align}
&  \Pr(j_1,j_2,\ldots j_{f+1} \vb n_1,n_2,\ldots n_f)= \notag\\
&M^{(n_f)}(j_{f+1},j_f)\cdots M^{(n_2)}(j_3,j_2)M^{(n_1)}(j_2,j_1)
\label{eqn3}
\end{align}
On the other hand, if the sequence of collision numbers is \emph{not} known,
the probability \emph{not} conditioned on this
information is given by
\begin{align}
&    \Pr(j_1,j_2,\ldots j_{f+1}) = \notag\\
&M^{(av)}(j_{f+1},j_f)\cdots M^{(av)}(j_3,j_2)M^{(av)}(j_2,j_1)
\label{eqn4}
\end{align}
where $M^{(av)}_{j'j}$ is the averaged Markov matrix,
\begin{equation}
  M^{(av)}(j',j) = \sum_n \Pr(n) M^{(n)}(j',j),
\label{eqn5}
\end{equation}
and $\Pr(n)$ the probability of $n$ collisions during a single time
interval. In the quantum case the single superoperator $\Tsup(\Dl t)$ is the
analog of an averaged Markov matrix, and it will allow us to correctly compute
the probability of a sequence of histories as long as we do not try and
condition it on more detailed information about the initial state of the
environment at the beginning of the time interval.

\subsection{Explicit form for the superoperator $\Tsup$}\label{sbct2.2}

As noted above, $\Tsup(\Dl t)$ only makes physical sense for $\Dl t$ greater
than some correlation time $\tau_c$.  Keeping this in mind, it is nonetheless
very convenient to think of the argument of $\Tsup(\Dl t)$ as a continuous
variable, which we shall hereafter denote by $t$, thus $\Tsup(\ta)$.  This
superoperator can be written in various ways, e.g., using Kraus operators or
as a matrix using some basis of the operator space of a qubit.  A convenient
basis is provided by the Pauli operators: $\sg_0=\id$, $\sg_1 = X$, $\sg_2 =
Y$, $\sg_3 = Z$, in terms of which we write
\begin{equation}
  \Tsup(\ta)\sg_j = \sum_k \Tmat_{kj}(\ta) \sg_k,
\label{eqn6}
\end{equation}
using a matrix $\Tmat$ of real coefficients whose first row (because $\Tsup$
is trace preserving) is $(1,0,0,0)$. The remaining rows constitute a
collection of 12 (real) parameters which are only constrained by inequalities
that ensure that $\Tsup$ is completely positive.  Applying $\Tsup(\ta)$ to a
density operator $\rho = \sum_j \bld{\rho}_j \sg_j$ at $t=0$, with the coefficients $\{\bld{\rho}_j\}$
regarded as a column vector $\bld{\rho}$, results in a density operator $\bar\rho = \sum_j
\bar{\bld{\rho}}_j \sg_j$ at time $t$, where $
\bar
{\bld{\rho}}
= \Tmat\cdot \bld{\rho} $.

Rather than explore the entire parameter space, we have assumed
that $\Tmat$ has the particularly simple form
\begin{equation}
\Tmat(\ta) = e^{\ta\Smat},\quad \Smat=  
\begin{pmatrix}
0 & 0 & 0 & 0\\
0 & 0 & -\omt & 0\\
0 & \omt & -2\gamma & 0\\
0 & 0 & 0 & -2\gamma
\end{pmatrix},
\label{eqn7}
\end{equation}
where $\omt$ is the precession frequency for $\ket{R}$ to $\ket{L}$ and back
again for the isolated molecule---the energy difference between $\ket{0}$ and
$\ket{1}$ is $\hbar\om$---and $\gm \geq 0$ is the \emph{rate of decoherence}. Justification based on scattering theory for this form of $\Tmat$ has been discussed in \cite{HornbergerEPL2007, PhysRevLett.103.023202}.

To see the motivation behind \eqref{eqn7}, first consider the case of the
isolated molecule with no decoherence, $\gm=0$.  Then 
\begin{equation}
\label{eqn8}
\Tmat(\ta)= \Rmat(\ta)=\begin{pmatrix}
1 & 0 & 0 & 0\\
0 & \cos \omt\ta & -\sin \omt\ta & 0\\
0 & \sin \omt\ta & \cos \omt\ta & 0\\
0 & 0 & 0 & 1
\end{pmatrix}
\end{equation}
corresponds to precession about the $z$ axis in a Bloch sphere picture.
Next, suppose that $\omt=0$, so that only decoherence is present.  Then
\begin{equation}
\Tmat(\ta) \approx \Dmat(\ta) =\begin{pmatrix}
1 & 0 & 0 & 0\\
0 & 1 &0 & 0\\
0 & 0 & 1-2\gm\ta & 0\\
0 & 0 & 0 &1-2\gm\ta
\end{pmatrix}
\label{eqn9}
\end{equation} 
when $\ta$ is small, and 
\begin{equation}\label{eqn10}
\Tmat(\ta)=\Dmat(\ta)\cdot \Rmat(\ta)+\Omat(\ta^2)
=\Rmat(\ta)\cdot \Dmat(\ta)+\Omat(\ta^2).
\end{equation}
with $\Omat(\ta^2)$ a second order correction. Thus
\eqref{eqn7} combines the competing effects of decoherence and the molecule's
internal dynamics.

The motivation behind \eqref{eqn9} is a simple physical picture in which
if the environment is initially in the state $\ket{E}$ its interaction with
the molecule during a collision corresponds to the unitary transformation
\begin{align} 
\ket{L}\ot\ket{E} &\rightarrow \ket{L}\ot\bigl( \sqrt{1-2\pp }\, \ket{E} +
\sqrt{2\pp }\, \ket{E'}\bigr),
\nonumber\\
\ket{R}\ot\ket{E} &\rightarrow \ket{R} \ot \bigl(\sqrt{1-2\pp }\, \ket{E} +
\sqrt{2\pp }\, \ket{E''}\bigr),
\label{eqn11} 
\end{align} 
where the environment states $\ket{E}$, $\ket{E'}$ and $\ket{E''}$ are
orthonormal. The intuitive idea is that the distinction between $\ket{L}$ and
$\ket{R}$ is carried off to the distinct environmental states $\ket{E'}$ and
$\ket{E''}$ with an amplitude that increases with $\pp $, a quantity lying
between 0 and 1/2 which is a measure of the effectiveness of the decoherence.
The decohering effect is unchanged if $\ket{E}$ on the right side of
\eqref{eqn11} is replaced with any other state $\ket{\bar E}$ as long as it is
orthogonal to $\ket{E'}$ and $\ket{E''}$, i.e., if the alteration does not
depend upon the difference between $\ket{L}$ and $\ket{R}$.
One can represent the channel corresponding to \eqref{eqn11} by three Kraus
operators associated with $\ket{E}$, $\ket{E'}$ and $\ket{E''}$, but an equally
good form uses just two Kraus operators $\sqrt{1-\pp }\;\id$ and $\sqrt{\pp
}\;X$ corresponding to a ``bit flip'' channel in
\cite{NielsenChuang:QuantumComputation} p.~376. When $\pp =0$ there is no
decoherence (a perfect channel) whereas for $\pp =1/2$ the collision
``collapses'' the molecule into either $\ket{R}$ or $\ket{L}$. If one sets $p=\gm\ta$ the superoperator corresponding to
the process \eqref{eqn11} is given by \eqref{eqn9}, and this makes sense for
$\ta$ of the order of the time between collisions. However, as noted above
in Sec.~\ref{sbct2.1}, the superoperator $\Tsup(\ta)$ can appropriately
represent a situation in which the number of collisions in the interval $\ta$
is a random quantity. 

The matrix $\Tmat(t)$ and the density operator $\bld{\rho}(t)$
thought of as a column vector satisfy the simple linear differential
equations:
\begin{equation}
  \frac{d\Tmat}{dt} = \Smat\cdot\Tmat,\quad
 \frac{d\bld{\rho}}{dt} = \Smat\cdot\bld{\rho}.
\label{eqn12}
\end{equation}
The second is equivalent to a master equation in Lindblad form
\begin{equation}
\label{eqn13}
\frac{d\rho}{dt} = -\ii[H ,\rho]/\hbar +\gamma (X\rho X-\rho),
\end{equation}
with $H /\hbar=\omt Z/2$ as in \eqref{eqn1}.

Solutions to \eqref{eqn12} and \eqref{eqn13} can of course be expressed as
linear combinations of exponentials of the form $e^{\lm_j t}$, where
\begin{equation}
  \lm_1 = 0,\; \lm_2 = -\gm + \xi,\; 
\lm_3 = -\gm -\xi,\; \lm_4 = -2\gm
\label{eqn14}
\end{equation}
\begin{equation}
\xi= \sqrt{\gm^2-{\omt}^2}
\label{eqn15}
\end{equation}
are the eigenvalues of the matrix $\Smat$. Note that $\lm_2$ and $\lm_3$ occur
in solutions of the form $e^{\lm t}$ to the damped oscillator equation
$d^2x/dt^2 + 2\gm dx/dt + {\omt}^2x=0$.  Thus for $\gm < {\omt}$ they are
complex conjugates of each other lying on a circle of radius ${\omt}$ in the
complex plane, corresponding to oscillatory solutions, while for $\gm >
{\omt}$ both are real and negative, corresponding to damped motion without
oscillation.  Critical damping $\gm={\omt}$ corresponds to a phase transition
in the sense of a changeover between two qualitatively different types of
behavior.  The explicit form of $\Tmat(t)$ is given in Appendix~\ref{apdxA}.

\xb
\section{General aspects of consistent histories and information\label{sct3}}
\xa

\xb
\subsection{Introduction to histories}
\label{sbct3.1}
\xa

In the (consistent or decoherent) histories formalism a \emph{history} is a
sequence of quantum properties, identified by projectors onto appropriate
subspaces of the quantum Hilbert space, at a succession of times $t_1 < t_2
<\cdots <t_f$; see Ch.~8 of \cite{\CQT}.  In the situation at hand we use a
sample space of mutually exclusive histories formed by assuming that at time
$t_m$ the properties of interest to us correspond to a collection
$\{P_m^{\al_m}\}$ of projectors which form a decomposition of the identity:
\begin{equation}
\sum_{\alpha_m}P_m^{\alpha_m}=\id, \quad
({P_m^{\alpha_m}})^\dagger={P_m^{\alpha_m}}=({P_m^{\alpha_m}})^2.
\label{eqn16}
\end{equation}
Here the subscript $m$ labels the time, while the superscript $\al_m$ is not
an exponent but instead a label to differentiate the projectors at this time.
Choosing at each time a property from the corresponding decomposition of the
identity yields a history represented by a projector 
\begin{equation}
  Y^{\vect{\al}} = P_1^{\al_1}\od P_2^{\al_2} \od \cdots P_f^{\al_f},\quad
\vect{\al} = (\al_1,\al_2,\ldots \al_f)
\label{eqn17}
\end{equation}
on the history Hilbert space $\breve \HC =\HC^{\od f} = \HC\od\HC\od \cdots
\HC$ formed by the tensor product of the Hilbert space with itself $f$ times.
Here $\od$ is a tensor product symbol with the same significance as $\ot$, but
employed to distinguish different times. The physical significance of
$Y^{\vect{\al}}$ can be seen by reading \eqref{eqn17} as ``property
$P_1^{\al_1}$ at time $t_1$ followed by property $ P_2^{\al_2}$ at time $t_2$
followed by\dots.''

For a closed system in which the unitary (Schr\"odinger) time development from
$t_m$ to $t_{m+1}$ is described by the operator $U_{m+1,m}$, probabilities
(probabilistic weights) can be assigned using the \emph{decoherence
  functional} \cite{RBGriffiths:ConsistentQuantumTheory}
\begin{align}
\Dfunc(Y^{\vect{\alpha}},Y^{\vect{\beta}}) = 
\Tr [& P^{\al_f}_{f} U_{f,{f-1}}\dotsm P^{\al_2}_{2}U_{2,1}P^{\al_1}_{1} \Psi_0 \notag\\
& P^{\bt_1}_{1}U_{1,2}
P^{\bt_2}_{2}\dotsm U_{{f-1},f}P^{\bt_f}_{f} ],
\label{eqn18}
\end{align}
where $\Psi_0$ is some initial state,
provided the \emph{consistency conditions}
\begin{equation}
  \Dfunc(Y^{\vect{\alpha}},Y^{\vect{\beta}}) = 0 
\text{ whenever } \vect{\al}\neq \vect{\bt}
\label{eqn19}
\end{equation}
are satisfied.  Here $\vect{\al}\neq\vect{\bt}$ means that for at least one
time $t_m$ it is the case that $\al_m\neq \bt_m$. When \eqref{eqn19} holds
one assigns the positive \emph{weight} $W(\vect{\al}) =
\Dfunc(Y^{\vect{\al}},Y^{\vect{\al}})$ to the history $Y^{\vect{\al}}$.  The
probability of each history is its weight divided by the total weight of all
the histories; if $\Psi_0$ is a normalized density operator this total weight
is 1 and the probability of history $\vect{\al}$ is $W(\vect{\al})$. 

The Hilbert space for the present discussion is $\HC = \HC_M\ot\HC_E$, where
$\HC_M$ is the Hilbert space of the molecule and $\HC_E$ that of the
environment.  However, the histories of interest to us refer to properties of
the molecule, not the environment, and we employ the usual convention that
$P_m^{\al_m}$ representing one of these properties can denote both a projector
on $\HC_M$ or its counterpart $P_m^{\al_m}\ot I_E$ on $\HC$.  For 
the initial state we let $\Psi_0 = I_M \ot \dya{\Phi_E}$, where 
\begin{equation}
  \ket{\Phi_E} = \ket{E_1}\ot\ket{E_2}\ot\cdots\ket{E_{f-1}}
\label{eqn20}
\end{equation}
is a ``giant'' tensor product state on the environment chosen in such a way
that during the time interval between $t_m$ and $t_{m+1}$ the molecule will
interact only with the piece $\ket{E_m}$ in this tensor product in a manner
determined by $U_{m+1,m}$; after that this part of the environment can be
ignored so far as the molecule is concerned.  In
particular, if we take a partial trace over the environment of the middle
portion on the right side of \eqref{eqn18} at time $t_2$, the interaction of
the molecule with $\ket{E_1}$ is chosen so that
\begin{equation}
  \Tr_E\left[ U_{2,1}P^{\al_1}_{1} \Psi_0 P^{\bt_1}_{1}U_{1,2} \right] =
\TC_{2,1}(P^{\al_1}_{1} P^{\bt_1}_{1}),
\label{eqn21}
\end{equation}
where $\TC_{2,1} = \TC(t_2-t_1)$ is the superoperator that maps the state of the
molecule at the beginning of this time interval to its state at the end.
In the same way, if the partial trace over the environment is carried out at
time $t_3$ the result will be 
\begin{equation}
  \TC_{3,2}(P_2^{\al_2} \TC_{2,1}(P^{\al_1}_{1} P^{\bt_1}_{1}) P_2^{\bt_2}),
\label{eqn22}
\end{equation}
with $ \TC_{3,2} = \TC(t_3-t_2)$, and similarly for later times.
Consequently, for our model the decoherence functional is given by
\begin{align}
\Dfunc(Y^{\vect{\alpha}},Y^{\vect{\beta}}) = \Tr_M [ &P^{\al_f}_{f}
  \Tsup_{f,f-1}(\dotsm P^{\al_2}_{2}\Tsup_{2,1}(P^{\al_1}_{1}\notag \\
 & P^{\bt_1}_{1})P^{\bt_2}_{2}\dotsm )P^{\bt_f}_{f} ],
\label{eqn23}
\end{align} 
an expression which no longer makes any (direct) reference to the environment.
See Sec.~III of \cite{PhysRevD.48.2728} for a more detailed argument.

If all the projectors in the decomposition $\{P_m^{\al_m}\}$ are rank 1, which
is to say they project onto pure states of the molecule, and the consistency
conditions are satisfied, then the probabilities (corresponding to the diagonal
elements of the decoherence functional \eqref{eqn23}) are those of a memoryless hopping process - a Markov
process.  If the time steps $\ta_m$ are identical and the same decomposition
is used at every time, this process is stationary (homogeneous, i.e.\ same
Markov matrix at each timestep), but in general it is nonstationary
(inhomogeneous).  Both cases are of interest for our model, as discussed
below in Sec.~\ref{sct4}.

\xa
\subsection{Forwards and backwards conditions}
\label{sbct3.2}
\xb

Finding collections of histories such that the consistency condition
\eqref{eqn19} is satisfied is made somewhat easier by the following
observation. Suppose it is the case that for every $m$ between 1 and $f-1$, if
$Q$ is a linear combination of the projectors in the set $\{P_m^{\al_m}\}$,
then $\TC_{m+1,m}(Q)$ is a linear combination of the projectors in the set
$\{P_{m+1}^{\al_{m+1}}\}$.  When this \emph{forward condition} is satisfied,
the family of histories will be consistent, as can be seen in the following
way.  The functional $\Dfunc(Y^{\vect{\alpha}},Y^{\vect{\beta}})$ in
\eqref{eqn23} will vanish if $\al_1\neq\bt_1$, since $P^{\al_1}_{1}
P^{\bt_1}_{1}=0$. If the forward condition is satisfied,
$\TC_{2,1}(P^{\al_1}_1)$ will be a linear combination of projectors in the
collection $\{P_2^{\al_2}\}$, and will therefore commute with any projector in
this collection.  Consequently, $P_2^{\al_2}\TC_{2,1}(P_1^{\al_1})P_2^{\bt_2}
= P_2^{\al_2}P_2^{\bt_2}\TC_{2,1}(P_1^{\al_1})$ will vanish whenever
$\al_2\neq\bt_2$ since $P_2^{\al_2}P_2^{\bt_2}=0$, and if it does not vanish
it will be some linear combination of the $\{P_2^{\al_2}\}$.  Proceeding in
the same way for larger $m$ one sees that
$\Dfunc(Y^{\vect{\alpha}},Y^{\vect{\beta}})$ will vanish if, for any $m$,
$\al_m\neq\bt_m$.

Note that the forward condition is a sufficient but not a necessary condition
for consistency.  The same is true of the \emph{backward condition}: for every
$m$ between $f$ and $2$ it is the case that if $Q$ is a linear combination of
the projectors in $\{P_m^{\al_m}\}$, then $\TC_{m,m-1}^\dagger(Q)$ is a linear
combination of the projectors in $\{P_{m-1}^{\al_{m-1}}\}$. Here
$\TC_{m,m-1}^\dagger$ denotes the adjoint of the superoperator with respect to
the Frobenius inner product: $\avg{\TC^\dagger(A),B} = \avg{A,\TC(B)}$
where $\avg{A,B} = \Tr(A\ad B)$.  The proof of consistency when the backward
condition is satisfied proceeds in the same way as for the forwards condition,
but in reverse.  Start with \eqref{eqn23} and rewrite the argument inside the
trace by first cycling $P_f^{\bt_f}$ to become the first term, and then
replacing $\TC_{f,f-1}$ with $\TC_{f,f-1}^\dagger$ acting on $P_f^{\bt_f}
P_f^{\al_f}$, and continue this cycling process to convert all $\TC$ to
$\TC^\dagger$. [Note that since $\TC$ is a (completely) positive superoperator, $(\TC\ad (A))\ad
= \TC\ad(A\ad )$.] In the case of qubits, the situation of primary interest for
the present paper, one can show that consistent families of histories of the
type \eqref{eqn17} must satisfy either the forward or the backward condition.

\xb
\subsection{Measuring information}\label{sbct3.3}
\xa

We will want to discuss and quantify the information about the initial state of the molecule as time goes on. We can do this within the context of the histories formalism. Alternatively, we can do this in the context of the quantum channel formalism, i.e., quantifying the distinguishability of density operators at the output of a quantum channel, and as we will see there is some connection between the two approaches.

Let us first consider information from the histories perspective. Suppose some
consistent family of histories uses the projective decompositions
$P_1=\{P_1^j\}$ at time $t_1$ and $P_m=\{P_m^k\}$ at time $t_m$, with $t_1<
t_m$ [for simplicity here we replaced the indices $\al_1$ and $\al_m$ in \eqref{eqn16} and \eqref{eqn17} with $j$ and $k$]. As in
\cite{PhysRevA.76.062320}, we will equate the notion of a projective
decomposition, like $P_1$, with a type of information about the system, in our
case the molecule. A convenient measure of how much of the $P_1$ type of
information about the molecule remains at time $t_m$ is the Shannon mutual
information
\begin{equation}
\label{eqn24}
H(P_1 \colo P_m)= H(P_1)+H(P_m)- H(P_1,P_m),
\end{equation}
where $H(P_1)$ is the familiar Shannon entropy. In particular if $P_1 $,
$P_m$, and $P_{m'}$ are projective decompositions associated with a consistent
family at three successive times, and if the probabilities correspond to a
Markov process, then (see, e.g., p.~510 of
\cite{NielsenChuang:QuantumComputation}) $H(P_1 \colo P_{m'})$ cannot be
greater than $H(P_1 \colo P_m)$: the information about the initial situation
can only decrease with time. For simplicity, in what follows we will set
$\Pr(P_1^j) = 1/d_1$ for all $j$, where $d_1$ is the number of projectors in
the decomposition $P_1$.  Then $H(P_1\colo P_1)=H(P_1)=\log d_1$, and hence
the information decays from its initial value of $\log d_1$ as time goes on.

Now, alternatively, consider the quantum channel perspective, where we will quantify how much of the $P_1$ type of information remains at time $t_m$ by measuring the \emph{distinguishability} of the conditional density operators at the output of the relevant quantum channel. (This approach was taken in \cite{PhysRevA.83.062338}.)  To measure distinguishability of density operators, in particular if these density operators do not commute, we need a measure that is inherently quantum-mechanical, which is provided by the Holevo function
\begin{equation}
\label{eqn25}
\chi(\{p_j,\rho_j\}):=S(\sum_j p_j\rho_j)-\sum_j p_jS(\rho_j)
\end{equation}
defined for an ensemble $\{p_j,\rho_j\}$, where $p_j$ is the probability
assigned to the density operator $\rho_j$, and $S(\rho):=-\Tr(\rho\log \rho)$
is the von Neumann entropy. Applying this measure to the ensemble $\{1/d_1,
\TC_{m,1} (P_1^j)\}$, where $\TC_{m,1} $ is the quantum channel that governs
the molecule's evolution from $t_1$ to $t_m$, gives a quantitative measure of
how much $P_1$ information remains at time $t_m$, and we write this as
\begin{equation}
 \hat\chi(P_1,\TC(t))=
 \hat\chi(P_1,\TC_{m,1}) : = 
 \chi( \{  \frac{1}{d_1} ,\frac{\TC_{m,1}(P_1^j)}{\Tr(P_1^j)} \}),
\label{eqn26}
\end{equation}
where $t=t_m-t_1$.

Equations~\eqref{eqn24} and \eqref{eqn26} give two alternative ways to measure the loss of information from the system over time. Equation~\eqref{eqn24} has the advantage of a clear conceptual interpretation, whereas Equation~\eqref{eqn26} has the advantage of being easy to compute since one does not need to go through the histories analysis to compute it. Fortunately, there is a connection between these approaches. It turns out, see the argument in Appendix~\ref{apdxC}, that for a family satisfying the forward consistency condition
\begin{equation}
\label{eqn27}
H(P_1 \colo P_m)= \hat\chi(P_1,\TC_{m,1}).
\end{equation}
A similar sort of connection holds for families satisfying the backward
consistency condition (but involving the adjoint channel $\TC_{m,1}\ad$), but
for simplicity we will focus on families satisfying the forward condition to
illustrate information flows in Sect.~\ref{sct5}.

One can also quantify information flow from the molecule to the environment with the quantum channel approach by using the complementary channel with superoperator $\TC^c$, introduced in Sec.~\ref{sbct2.1}. In fact, $\TC^c$ is completely determined by $\TC$ up to an isometry on its output (the environment), which does not affect distinguishability measures like $\chi$. Hence, the following information measure is well-defined:
\begin{equation}
 \hat\chi(P_1,\TC^c(t))=
  \hat\chi(P_1,\TC^c_{m,1}) : = \chi( \{  \frac{1}{d_1} ,\frac{\TC^c_{m,1}(P_1^j)}{\Tr(P_1^j)} \}),
\label{eqn28}
\end{equation}
where $t=t_m-t_1$.  It quantifies the amount of the $P_1$ type of information
about the molecule (at time $t_1$) that is present in the environment at time
$t_m$. Though one cannot in general equate this $\hat\chi(P_1,\TC^c_{m,1}) $
with a Shannon mutual information between the molecule and the environment,
the former provides, as is well-known (e.g., p.~531 of
\cite{NielsenChuang:QuantumComputation}), an upper bound on the latter.

We note that there can be a tradeoff in sending information to the environment and preserving it in the molecule, which is most dramatic for complementary or \textit{mutually-unbiased} bases $P_1$ and $P'_1$:
\begin{align}
\label{eqn29}
\hat\chi(P_1,\TC_{m,1})+ \hat\chi(P'_1,\TC^c_{m,1})\leq \log d_1.
\end{align}
This inequality is from Corollary 6 of \cite{PhysRevA.83.062338}.

\xb
\section{Consistent families for our model}\label{sct4}
\xa

\xb
\subsection{Differential equations}\label{sbct4.1}
\xa

Our model has only two states, and therefore any (nontrivial) decomposition of
the identity involves only projectors of rank 1 onto pure states. Thus a
consistent history family corresponds to a two-state Markov process
(sometimes called a ``telegraph process''); in general this process is
nonstationary:\ the transition rates depend upon the time.  While such a
process can be discussed using discrete times separated by finite intervals,
the results are simpler and the mathematical expressions more transparent if
one adopts a continuous time approximation with differential equations in
place of difference equations.  It should, of course, be kept in mind that the
processes here described are not truly continuous, since time intervals
shorter than the correlation time $\tau_c$ introduced in Sec.~\ref{sct2} lack
physical significance.  The continuous time approach should be satisfactory as
long as both $\omt\tau_c$ and $\gm\tau_c$ are small compared to 1.  Note that
this condition can still be true even when $\gm$ is large, as long as $\tau_c$
decreases as $1/\gm$, which seems physically plausible.

For families satisfying the forward consistency condition the relevant
differential equations can be obtained in the following way. At a particular
time the decomposition of the identity will correspond to two projectors, call
them $\rho_0$ and $\rho_1$, represented by end points or antipodes of a
diameter of the Bloch sphere.  Let the direction of this diameter be denoted
by the usual polar and azimuthal angles $\th$ and $\phi$: the $z$ axis at
$\th=0$ and the $x$ axis at $\th=\pi/2$, $\phi=0$. Which end of the diameter
corresponds to these angles does not matter for the following discussion.  The
locations of these end points after a short time interval is determined by the
master equation \eqref{eqn13}.  One can show that because of the form of
$\Smat$ in \eqref{eqn7} they are still located on a diameter of the Bloch
sphere, but are now a bit closer to its center. The rate of change of the diameter's direction is represented by the differential equations
\begin{equation}
\label{eqn30}
\frac{d\phi}{dt}=\omt-\gm \sin 2\phi,\quad
\frac{d\th}{dt}=\gm \sin 2\th \cos^2 \phi,
\end{equation}
whereas the shift towards the center can be used to calculate the instantaneous
transition rate
\begin{equation}
\kappa = \gm(1-  \sin^2 \th \cos^2 \phi),
\label{eqn31}
\end{equation}
which enters the rate equations
\begin{equation}
  dp_0/dt = \kp (-p_0 + p_1),\quad  dp_1/dt = \kp (p_0 - p_1)
\label{eqn32}
\end{equation}
for the probabilities associated with these two states. The backwards
consistency condition can be analyzed in a similar way, and leads to the
differential equations
\begin{equation}
\frac{d\phi}{dt}=\omt+\gm \sin 2\phi,\quad
\frac{d\th}{dt}=-\gm \sin 2\th \cos^2 \phi.
\label{eqn33}
\end{equation}
governing the direction of the diameter, and to exactly the same expression
\eqref{eqn31} for the transition rate.  For a more detailed derivation of
these formulas see Appendix~\ref{apdxBB}.

\xb
\subsection{Stationary families}\label{sbct4.2}
\xa

If the angles $\th$ and $\phi$ which determine the diameter for the projectors
forming a consistent family do not change with time the Markov process is
stationary or homogeneous, in the sense that the states and the transition
probabilities do not change with time; of course the actual state of the
molecule is varying randomly as it hops back and forth between the two states.
The simplest case is what we call the $z$ family, in which $\th=0$ (or $\pi$),
thus $d\th/dt=0$ in \eqref{eqn30} or \eqref{eqn33} and $d\phi/dt$ is
irrelevant.  The two projectors $(\id+\sg_z)/2$ and $(\id-\sg_z)/2$ correspond,
respectively to the even parity (ground) and odd parity (excited) states of
the isolated molecule.  Thus we have a two-state stationary Markov process in
which the molecule spends a certain amount of time in the even parity state
before flipping instantaneously (in our continuous time approximation) to the
odd parity state where it remains for a random time interval before flipping
back.  The time $\tau$ between flips is a random variable with an exponential
distribution $e^{-\gm \tau}$, since setting $\th=0$ in \eqref{eqn31} gives
\begin{equation}
 \kp_z = \gm
\label{eqn34}
\end{equation}
for the transition rate. On average the molecule spends an equal
amount of time in both states, which means that in our model the environment
has an effective temperature $T\gg \hbar\omt/k_B$.

\begin{figure}[h]
\centering
\includegraphics[scale=0.8]{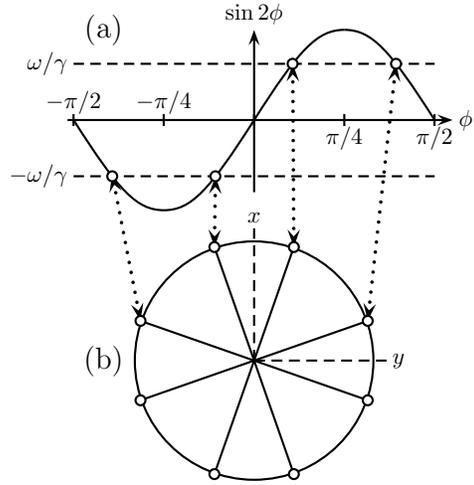}
\caption{(a) The steady-state solutions for $\phi$ correspond to the
  intersections of $\sin 2\phi $ (solid curve) with $\pm \omt/\gm$ (dashed
  lines), shown here for $\gm > \omt > 0$ (``strong decoherence'' regime). (b)
  These steady-state solutions are plotted schematically on the Bloch sphere,
  as if the $z$-axis is going into the page. \label{fgr1}}
\end{figure}

In addition to the $z$ family just discussed there are stationary
families in which the projectors correspond to points in the $x$-$y$ or
equatorial plane of the Bloch sphere, so $\th=\pi/2$ with $d\phi/dt=0$
in \eqref{eqn30} and \eqref{eqn33}, and thus
\begin{equation}
  \sin 2\phi = \pm (\omt/\gm).
\label{eqn35}
\end{equation}
For $0<\omt/\gm<1$ there are four solutions as shown in Fig.~\ref{fgr1}, which
coalesce into two for $\omt/\gm=1$. For $\omt/\gm > 1$ these families
disappear, leaving the $z$ family as the only stationary family.  In the limit
of strong decoherence, small $\omt/\gm$, two of the families approach the $x$
axis and two the $y$ axis of the Bloch sphere, so we shall refer to them as
the dressed $x$- and dressed $y$-families.  The associated transition rates
$\kp_x$ and $\kp_y$ are given by $-1/2$ times the corresponding eigenvalues of
$\Smat$, see \eqref{eqn14}:
\begin{equation}
  \kp_x = -\lm_2/2 = (\gm-\xi)/2,\quad 
\kp_y = -\lm_3/2 = (\gm+\xi)/2. 
\label{eqn36}
\end{equation}

As $\gm/\omt$ becomes very large the dressed $x$-families approach the $x$ or
chirality basis $\ket{R}$ and $\ket{L}$ of \eqref{eqn2}, and the transition
rate $\kp_x\approx\omt^2/4\gm$ becomes very slow.  Thus these families
represent long-lived (almost) chiral states when decoherence is rapid compared
with the the tunneling rate. (But see the further discussion in
Sec.~\ref{sbct4.5}.)

\xb
\subsection{Nonstationary families}\label{sbct4.3}
\xa

\begin{figure*}[ht]
\centering
\includegraphics{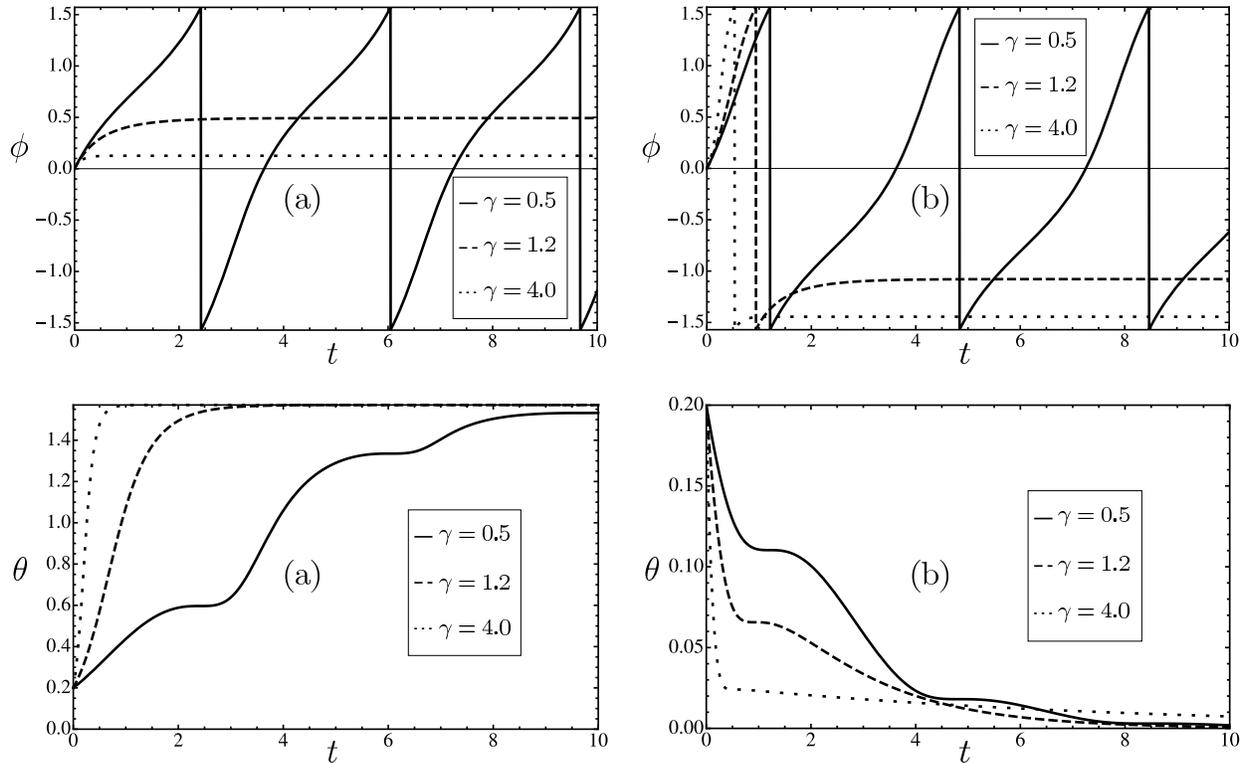}
\caption{Time evolution of the Bloch-sphere angles $\phi$ and $\th$ for the
  consistent description associated with the (a) forward and (b) backward
  conditions. In this case, $\phi(0)=0$, $\th(0)=0.2$, $\om=1$, and
  $\gm=0.5,1.2,4$ respectively for the solid, dashed, and dotted curves. The angles $\phi$ and $\th$ are in units of radians, and the time $t$ is in units of $1/\om$. \label{fgr2}}
\end{figure*}

The equations \eqref{eqn30} and \eqref{eqn33} can be integrated in closed
form to obtain the bases corresponding to nonstationary consistent families,
Appendix~\ref{apdxB}.  However, the solutions are fairly complicated
expressions.  The time evolution for some cases in which $\omt=1$ and $\gm$ is
either less than or larger than $1$ is shown in Fig.~\ref{fgr2}.
For $\gm < \omt$ the diameter rotates continuously about the $z$ axis (the
discontinuities in $\phi$ are of course artifacts of the plot) 
with an angular frequency of
\begin{equation}
  \eta = \sqrt{\omt^2 - \gm^2},
\label{eqn37}
\end{equation}
while the polar angle $\th$ tends either to $\pi/2$ for the forward or to $0$
(equivalently, $\pi$) for the backward consistency condition. In the limit in
which $\gm$ tends to 0, no decoherence, one has a simple rotation of the
diameter of the consistent family about the $z$ axis at a rate $\omt$ with
$\th$ fixed. This same tendency is seen in the dependence of $\th$ on time
when $\gm > \omt$, whereas $\phi$ more or less rapidly
approaches one of the values corresponding to a stationary family.  Note that
along with the continuous change of basis there is a random flipping from
one of the basis states to the other at a rate given by \eqref{eqn31}, so one
is dealing with a nonstationary Markov process.

Thus in the Bloch sphere picture, for $\gm < \omt$, the families ``corkscrew''
about the $z$ axis (going away from or towards this axis for the forward or
backward families, respectively), with $\phi$ periodically coming back to the
same value at time intervals that are integer multiples of $\pi / \eta$.  If
$\th = \pi /2$ it remains constant, so the same basis reoccurs after an
interval of $\pi / \eta$.  If only these discrete times are considered, the
result is what one might call a \emph{stroboscopic} family which can be
thought of as a discrete time stationary Markov process.

\xb
\subsection{Phase transition}
\label{sbct4.4}
\xa

As the parameters $\gm$ and $\omt$ vary there is a phase transition, a
qualitative change of behavior, when they are equal.  This manifests itself in
a variety of related ways.  For $\gm < \omt$ the eigenvalues of $\Smat$
include a complex-conjugate pair $\lm_2$ and $\lm_3$, \eqref{eqn14}, which
coalesce into a single degenerate eigenvalue at the transition, and
thereafter, for $\gm > \omt$, become a pair of distinct real eigenvalues.
This is, of course, precisely the behavior one finds in a classical
one-dimensional oscillator when the damping passes through the critical
value. For $\gm > \omt$ these eigenvalues are the decay rates for the
dressed-$x$ and dressed-$y$ continuous stationary Markov processes discussed in
Sec.~\ref{sbct4.2}.  On the other hand, as $\gm$ decreases towards $\omt$ from
above, the four stationary families shown in Fig.~\ref{fgr1}(b) coalesce into
two, corresponding to diameters of the Bloch sphere midway between the $x$ and
$y$ axes, and for $\gm < \omt$ they no longer exist: the only remaining
continuous stationary family is the $z$ family.  As noted above in
Sec.~\ref{sbct4.3}, for $\gm < \omt$ there is a new class of ``stroboscopic''
families defined using a periodic time interval.  As $\gm$ approaches $\omt$
from below this period becomes infinitely long.  The behavior of nonstationary
continuous families is also different for $\gm < \omt$ and $\gm > \omt$.  For
the former $\phi$ increases indefinitely and monotonically with time, although
this motion, which is simply linear when $\gm=0$, becomes more and more
``jerky'' as $\gm$ increases towards $\omt$. See the example for $\gm/\omt =
1/2$ in Fig.~\ref{fgr2}.  For $\gm > \omt$, $\phi$ approaches a fixed value
with increasing time, and no longer ``winds.'' (Again, the damped harmonic
oscillator provides a helpful analogy.) One might say that the nonstationary
continuous families transition from a damped oscillatory character to a purely
damped character as $\gm/\omt$ increases, passing through the critical value
of 1.

In terms of its mathematical structure as represented in the master equation
this is a dynamical quantum phase transition of the sort discussed in quantum
optics for two level systems in Ch.~11 of \cite{WlMl94} and in \cite{GrLs10},
and in a more general context in \cite{Pstw07,Rttr09,Rttr10}.  It appears that
the vanishing of the inversion transition in ammonia is of this type; see the
discussion in Sec.~\ref{sbct6.2} below. We believe that ours is the first
attempt to explore dynamical properties near such a transition using the
histories approach.

\xb
\subsection{Physical interpretation}
\label{sbct4.5}
\xa

Each consistent family contains a collection of histories, and each history a
particular succession of micrscopic properties (subspaces of $\HC_M$).  One
and only one history from this collection will describe the behavior of a
particular molecule during a particular interval of time.  There is no need to
make any reference to measurements, though it is in principle possible (i.e.,
does not violate the laws of quantum mechanics) to use a succession of
suitably idealized measurements to determine which of these histories is
actually realized.  But because one is dealing with a system exhibiting
``quantum'' behavior, i.e., in a regime in which a classical description is
not adequate, it is important to keep in mind certain respects in which
quantum descriptions differ from their classical counterparts.

In particular, two consistent families of histories may be mutually
incompatible with each other in such a way that they cannot be combined into a
single description that makes sense. A well-known example is a spin-half
particle where incompatibility arises from the fact that the operators for
angular momentum in different directions do not commute with each other, and
hence have no common eigenvectors.  It makes good (quantum) sense to say, for
example, that $S_x=+1/2$ (in units of $\hbar$), or that $S_z=+1/2$, but there
is no quantum property, no subspace in the Hilbert space, that corresponds to
$S_x=+1/2$ \emph{and} $S_z=+1/2$. So one cannot ascribe simultaneous existence
to $S_z$ and $S_x$.  In the consistent histories approach this inability to
combine incompatible descriptions is codified as the single framework rule,
and the consistency conditions discussed above in Sec.~\ref{sct3} serve to
extend this rule from a single time to a sequence of times. In addition, just as two incompatible consistent families or descriptions
cannot be combined into a single description, they also cannot be compared: it
makes no sense to ask which of two incompatible families is the ``correct''
one, or to look for some law of nature that single out one against another.
Each consistent family provides its own quantum description in a way roughly
analogous to looking at a mountain from different locations.  For a detailed
discussion of these points we refer the reader to \cite{Grff12}.

With reference to a tunneling molecule, consider the situation in which $\gm$
is much larger than $\omt$, strong decoherence.  There is a stationary family,
Sec.~\ref{sbct4.2}, the ``chiral'' family, in which the molecule hops back and
forth at a comparatively slow rate between the (dressed) left-handed and
right-handed chiral states. (There are actually two of these dressed-$x$
families, but when decoherence is strong there is very little difference
between them.)  This is the family to use if one is interested in
understanding why a specific chirality, the left or right-handed form of the
molecule, can persist for a very long time in a situation of strong
decoherence.  It provides a description in terms of a stochastic two-state
Markov process in which the rate of hopping from left to right-handed or vice
versa is a well-defined function, $\kp_x$ in \eqref{eqn36}, of the parameters
that enter the model.  (Each hop is instantaneous on the time scale used for
our description, in which intervals less than the correlation time $\tau_c$ do
not enter; see Sec.~\ref{sct2}.)
As the rate of decoherence decreases, the hopping time becomes shorter and the
amount of ``dressing'' required to produce a consistent family increases,
which means that even though this family continues to provide a correct
quantum description, it no longer corresponds to a simple physical picture of
a definite left- or right-handed molecule when $\gm$ becomes comparable to
$\om$. 

In addition to this chiral family there is a $z$ or ``parity'' family in which
the molecule hops back and forth at random time intervals between parity
eigenstates (energy eigenstates of the isolated molecule), at a rate given by
the decoherence rate $\gm$, see \eqref{eqn34}.  The parity family is
incompatible with the chiral family discussed above, and they cannot be
combined.  One should not try and imagine them as going on simultaneously; to
do so would be to make the same mistake as supposing that $S_x$ and $S_z$ for
a spin-half particle can simultaneously possess values.  On the other hand,
just as it is possible to measure either $S_x$ or $S_z$, but not both
simultaneously, it is also possible in principle (without violating the laws
of quantum mechanics) to determine by measurements the succession of events
that occur in a parity family, or by a different set of measurements those
occurring in a chiral family.  Thus a relatively rapid but random flipping
back and forth between parity eigenstates is a valid physical picture of the
succession of microscopic states of the molecule, one which can be used both
when the decoherence is strong and when it is weak.  There is in addition a
third stationary family for $\gm > \om$, the dressed $y$ family, which has a
relatively rapid hopping rate in the strong decoherence regime, and eventually
merges with the chiral family as the decoherence rate decreases.  We do not
have a simple name or physical interpretation for this family.

In the regime where decoherence is weak, $\gm <\om$, there are no truly
stationary families, apart from the parity family discussed above.  A
relatively simple nonstationary family is the one that employs an
``equatorial'' basis in the $x$-$y$ plane of the Block sphere, $\th=\pi/2$,
rotating at an average angular speed $\eta$, see \eqref{eqn37}. Let us call
this the ``tunneling'' family, since it corresponds in physical terms to the
molecule oscillating back and forth between the two potential wells.  As $\gm$
increases the rate of tunneling decreases and eventually goes to zero at the
phase transition $\gm=\om$.  In addition to the tunneling, the phase $\phi$
undergoes random changes by $\pi$, instantaneous on the time scale we are
using, at a rate, \eqref{eqn31}, proportional to $\gm$, but also depending on
the value of $\phi$. Thus we have a nonstationary Markov process.  The random
flipping rate increases with $\gm$ at the same time as the tunneling rate is
decreasing, so the simple physical picture of the molecule tunneling from one
potential well to the other breaks down upon approaching the phase transition
$\gm=\om$. For larger values of $\gm$ this consistent family no longer exists.

\xb
\section{Information Flows\label{sct5}}
\xa

In the previous section we found various consistent frameworks for discussing
the stochastic trajectory (for our model) of a tunneling molecule. We now wish
to study the dynamics of information, e.g., the loss of information about the
molecule's original state as time progresses. Section~\ref{sbct3.3} discussed
how the Shannon mutual information between the original state and the state at
some later time, for the forward or backward consistent family, is equivalent
to a particular Holevo $\chi$ quantity. Here, for simplicity, we will focus on
families satisfying the forward condition, for which the information remaining
in the molecule is given by \eqref{eqn27}, and that flowing to the environment
by \eqref{eqn28}.

\begin{figure}[t]
\centering
\includegraphics[scale=0.99]{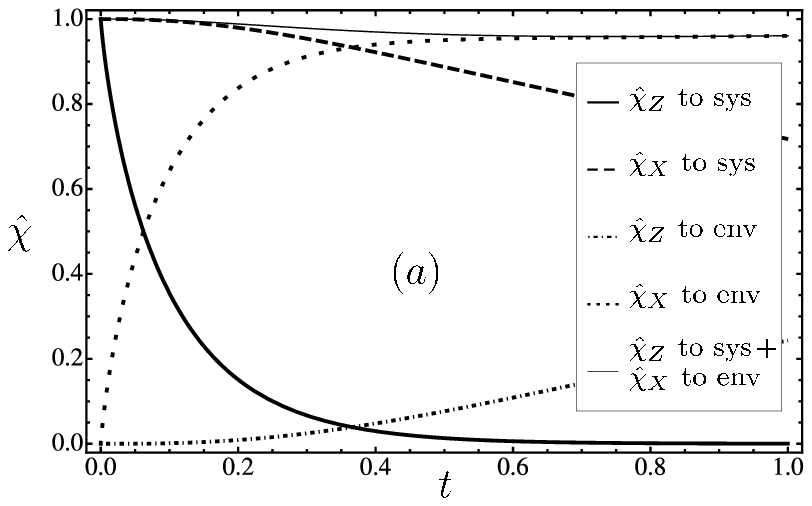}
\includegraphics[scale=0.99]{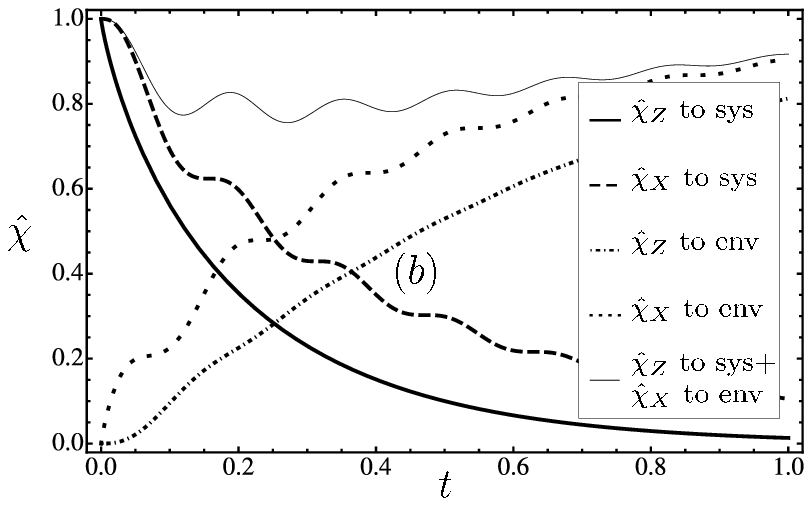}
\caption{Information flows in terms of the $\hat \chi$ information measure (see text for definition), (a) in the strong decoherence regime with $\gm = 2$ and $\omt = 0.8$, and (b) in the weak decoherence regime with $\gm = 1$ and $\omt = 20$. The measure $\hat \chi$ is in units of bits, and the time $t$ is in units of $0.8 / \om$ in (a) and $20/\om$ in (b).}
\label{fgr3}
\end{figure}

These quantities are shown in Fig.~\ref{fgr3} for $P_1 = Z$ (parity basis) and
$P_1=X$ (chirality basis), both for information remaining in the molecule
$\TC(t)$ and that flowing to the environment $\TC^c(t)$. Figure~\ref{fgr3}(a)
shows a case of strong decoherence, $\gm/\om =2.5$, where the curves as a
function of time are quite smooth, consistent with the fact,
Fig.~\ref{fgr2}(a), that the consistent family is rapidly approaching a
stationary family. For weak decoherence, Fig.~\ref{fgr3}(b) with
$\gm/\om=0.05$, the consistent family is not stationary and the alternating
rises and plateaus reflect this fact.  The top curves in both (a) and (b)
represent the sums, see \eqref{eqn39} below, for one type of information
remaining in the molecule and a mutually unbiased type flowing to the
environment.

For short times the individual information measures can be computed, using
the expressions for $\TC$ and $\TC^c$ in Appendix~\ref{apdxA} and noting
that $S(\TC^c(\dya{\psi}))=S(\TC(\dya{\psi}))$ for any pure state
$\ket{\psi}$, to obtain
\begin{align}
\label{eqn38}
\hat{\chi}(X,\TC(\ta))&=1 - O[\ta^2\log (1/t)],\notag \\
\hat{\chi}(Z,\TC(\ta))&=1 - \gm t\log(1/\gm t)-\gm t + O[\ta^2\log (1/t)],\notag \\
\hat{\chi}(X,\TC^c(\ta))&= \gm t\log(1/\gm t)+\gm t + O[\ta^2\log (1/t)],\notag \\
\hat{\chi}(Z,\TC^c(\ta))&= O[\ta^2\log (1/t)],
\end{align}
Here $O[\;]$ means that the correction term is of this or possibly some
higher order.  These expressions are consistent with $\hat{\chi}(X,\TC(\ta))$
and $\hat{\chi}(Z,\TC^c(\ta))$ having zero slope at $t=0$, and
$\hat{\chi}(Z,\TC(\ta))$ and $\hat{\chi}(X,\TC^c(\ta))$ having infinite slope
at $t=0$, as depicted in Fig.~\ref{fgr3}. That $\hat{\chi}(Z,\TC^c(\ta))$ in
\eqref{eqn38} has no term linear in $t$ seems plausible in that the
decoherence mechanism in our model has been chosen specifically to carry $X$
information into the environment. The uppermost curves in Figs.~\ref{fgr3}(a) and \ref{fgr3}(b) represent
\begin{align}
\label{eqn39}
\hat{\chi}(Z,\TC(\ta))+\hat{\chi}(X,\TC^c(\ta))&= \notag\\
\hat{\chi}(Z,\TC^c(\ta))+\hat{\chi}(X,\TC(\ta))& =1 - O[\ta^2\log (1/t)],
\end{align}
where the first equality comes from Theorem~3 of \cite{PhysRevA.83.062338},
and the second  from \eqref{eqn38}; the correction must be negative 
in view of the bound in \eqref{eqn29}.

Another consequence of \eqref{eqn38} is that for small $\ta$ the flow of chiral
information to the environment is compensated by a decrease of parity
information remaining in the molecule:
\begin{equation}
\frac{\mathrm{d}}{\mathrm{d}\ta}\hat{\chi}(X,\TC^c(\ta))
 =-\frac{\mathrm{d}}{\mathrm{d}\ta}\hat{\chi}(Z,\TC(\ta)) + O[\ta \log (1/t)].
\label{eqn40}
\end{equation}
However, as noted above, both sides of \eqref{eqn40} diverge
logarithmically as $\ta \to 0$.  Of course, these expressions lack physical meaning for times
shorter than $\tau_c$, and thus the divergence is a mathematical artifact.
Nonetheless, this makes it difficult to define rates of flow of information in
a mathematically clean way using the $\hat{\chi}$ measure.  

An alternative which avoids the divergence is to replace the von Neumann
entropy $S$ in the definition \eqref{eqn25} with the quadratic entropy
\begin{equation}
\label{eqn41}
S_Q(\rho)=1-\Tr(\rho^2).
\end{equation}
In the case of a qubit channel with $W$ an orthonormal basis, projectors
$W^1+W^2=I$, and Pauli operator $\sg_W := W^1-W^2$, the measure defined in \eqref{eqn26} becomes
\begin{equation}\label{eqn42}
\hat{\chi}_Q\left(W,\TC(\ta)\right)=
\frac{1}{2}\Tr\left[\Bigl(\TC(\sg_W)\Bigr)^2 \right],
\end{equation}
with a similar expression for the complementary channel if $\TC$ is
replaced by $\TC^c$. One can use the usual Pauli representation to write
\begin{equation}
\label{eqn43}
\sg_W = \vect{n}\cdot{\vect\sigma},
\end{equation} 
where the $x$, $y$, and $z$ components of the $\sg_W$ Pauli operator are given by the unit vector
\begin{equation}
  \vect{n}=\{n_x,n_y,n_z\}
=\{\sin \theta \cos \phi,\sin \theta \sin \phi, \cos \theta\}.
\label{eqn44}
\end{equation}
Using the short-time expressions for $\TC$ and $\TC^c$ given in Appendix~\ref{apdxA}, one finds that for the direct and complementary channels
\begin{align}
\hat{\chi}_Q\left(W,\TC(\ta)\right)
&=1-4\gamma\ta\left[1-n_x^2 \right] + O(\ta^2),
\label{eqn45}\\
\hat{\chi}_Q\left(W,\TC^c(\ta)\right)
&=4\gamma\ta n_x^2 + O(\ta^2).
\label{eqn46}
\end{align}

Setting $W=X$ in \eqref{eqn46} and $W=V$ in
\eqref{eqn45}, where $V$ is some basis in the $y$-$z$  plane and
thus mutually unbiased relative to $X$, the analog of \eqref{eqn40} with
$\hat{\chi}$  replaced by $\hat{\chi}_Q$ is 
\begin{equation}
\frac{d}{d\ta} \hat{\chi}_Q(X,\TC^c(\ta)) = -\frac{d}{d\ta}\hat{\chi}_Q(V,\TC(\ta)) =4\gamma
\label{eqn47}
\end{equation}
at $t=0$, so the derivatives are now finite. 
Thus when one uses the $\hat{\chi}_Q$ measure the rate of flow of $X$
information to the environment equals the rate of decrease within the molecule
of any type of information associated with a basis in the $y$-$z$ plane.

\xb
\section{D$_2$S$_2$ and NH$_3$}
\label{sct6}
\xa

\subsection{D$_2$S$_2$}
\label{sbct6.1}

An order-of-magnitude calculation of the decoherence rate $\gm$ for a
D$_2$S$_2$ molecule immersed in a gas of helium atoms shows that this system
is in the strong decoherence regime under typical conditions.  The flux $q$ of
helium atoms (atoms per unit area per unit time) is their concentration times their average velocity, given by $q= P\cdot\sqrt{8/(\pi
  m_{\text{He}} k_b T)}$, assuming an ideal gas with pressure $P$ and
temperature $T$, with $m_{\text{He}}$ and $k_b$ the mass of a helium atom and
Boltzmann's constant \cite{KauzmannKineticTheory}.  At room temperature $T=300$ K and a pressure $P$ of 1
atmosphere this gives $q\approx 3 \times 10^{28}$ atoms s$^{-1}$ m$^{-2}$.
Multiplying $q$ by approximate cross sections $\sigma_{\text{col}}\approx 1000
a_0^2$ and $\sigma_{\text{dec}}\approx 100 a_0^2$ for collisions and
decoherence, taken from \cite{PhysRevLett.103.023202}, with $a_0$ the Bohr
radius, leads to a collision rate of approximately $9 \times 10^{10} s^{-1}$ and a decoherence rate of
\begin{equation}
\gamma\approx 9 \times 10^{9} s^{-1}.
\label{eqn48}
\end{equation}
The estimated tunneling rate is $\omt\approx 176$ rad/s after correcting\footnote{Private
  communication from K. Hornberger} the value published in
\cite{PhysRevLett.103.023202} by a factor of $2\pi$. Thus
\begin{equation}\label{eqn49}
\gamma/\omt\approx 5 \times 10^7,
\end{equation}
which means strong decoherence, for which 
our results in Sec.~\ref{sct4} indicate that chirality is
both a consistent description as well as a stable property, to a very good
approximation.  Probing the regime $\gm \leq \om$ for this 
molecule would seem quite difficult as it would involve very low pressures. Replacing deuterium with hydrogen and/or sulfur with oxygen leads to chiral molecules, e.g.\ H$_2$O$_2$, that have significantly higher tunneling frequencies \cite{Quack2001} and hence may be candidates for probing the $\gm \leq \om$ regime.

\subsection{NH$_3$}
\label{sbct6.2}

In the electronic ground state of ammonia NH$_3$ the nitrogen lies to one side
of the plane defined by the three hydrogens, but there is a relatively low
potential barrier separating it from the mirror image state on the other side,
and tunneling in this double-well potential exhibits itself in the well-known
inversion transition at a frequency of about 24 GHz.  Although the molecule is
not chiral, when it is rotating about an axis passing through the nitrogen and
the midpoint between the hydrogen atoms the symmetry operation of parity
(inversion) moves the nitrogen to the other side of the plane, changing the
sign of the electric dipole, while leaving the angular momentum unchanged.
Consequently, the energy levels with a nonzero quantum number $K$ for this
component of angular momentum are split into two parity eigenstates by the
inversion transition in a way similar to that in a chiral molecule.

The tunneling transition has been observed directly by microwave absorption,
which at low pressure exhibits a set of closely-spaced lines associated with
the different rotational states \cite{Twns46}.  As the pressure increases the
lines broaden and merge, and the center of the merged line shifts towards
lower frequencies, reaching zero frequency at a pressure of about 2
atmospheres \cite{BlLb50}.  It has been suggested, e.g. \cite{Wght95}, that at
pressures above this transition the ammonia molecule adopts a ``pyramidal''
shape with the nitrogen on one side of the hydrogen plane, analogous to the
shape of a chiral molecule with a definite handedness. Deuterated ammonia
ND$_3$ shows similar behavior, except that the low pressure tunneling
frequency is now at 1.6 GHz, and the center of the broadened line tends to
zero frequency at a pressure of 0.12 atmospheres \cite{BrMr53}.

The shift towards zero frequency has been analyzed theoretically using two
different approaches.  The first, exemplified by \cite{JLPT02}, and with
similar ideas in \cite{GrSc04,GnBr11} among others, starts with the
observation that since the ammonia molecule possesses a significant electric
dipole moment when the nitrogen is on one side of the hydrogen plane there
will be a strong dipole-dipole interaction between nearby molecules.  It is
then proposed that this produces a sort of mean-field effect in which the
polarization of one molecule influences its neighbors in such a way that
eventually as the pressure increases and the molecules come closer together,
the double well potential for a single molecule is changed into one with a
single minimum on one side of the hydrogen plane, resulting in molecules of
pyramidal shape.

An alternative approach found in \cite{BRvn65,BRvn66,HrDn07,BhBs11} focuses instead
on the decohering effects of collisions between gas molecules. It is argued
that these collisions in addition to broadening the lines can also lower the
tunneling frequency as the pressure, and thus the collision rate, increases.
From this perspective the electric dipole-dipole interaction, while
significant in determining the collision cross section and the effects of
collisions, is not the fundamental source of the line shift to lower
frequencies. The latter ought still to be present if ammonia is a dilute
component in a nonpolar buffer gas. Of particular significance for this second point of view is the work of
Ben-Reuven \cite{BRvn65,BRvn66}, who argued on theoretical grounds that when
proper account is taken of the effects of collisions the line shape,
absorption as a function of frequency, is not adequately represented by the
Van Vleck and Weisskopf formula \cite{VVWs45} used earlier in \cite{BlLb50} to
analyze the experimental data.  He proposed an alternative line shape function
with three parameters, $\gm$, $\zt$ and $\dl$, proportional to the collision
rate, and thus the pressure, to fit the experimental microwave absorption data
for NH$_3$ and (with a different choice of parameters) ND$_3$ over a range of
frequencies and pressures sufficient to include that at which the tunneling
frequency goes to zero.  It is noteworth that this fit was achieved for all
pressures and frequencies using just these three parameters, whereas the
earlier analysis of Bleaney and Loubser \cite{BlLb50} was carried out by
adjusting two parameters separately for each pressure.

The validity of Ben-Reuven's analysis is supported by the fact that more
recent data on microwave absorption by ammonia in mixtures of hydrogen and
helium (of interest in studies of the atmospheres of Jupiter and the other
giant planets) has been fitted using his line shape formula for the tunneling
transition \cite{DvSK11} with, of course, different choices of parameters for
the different species scattering from the ammonia molecule.  Since neither
hydrogen nor helium has an electric dipole moment, this tends to support the
idea that collisions, rather than dipole-dipole interactions as such, are what
drive the transtion frequency to zero in pure ammonia gas as the pressure
rises.  The numbers given in \cite{DvSK11} would suggest a phase transition at
about 20 atmospheres for ammonia in a buffer gas of hydrogen at room
temperature.  Replacing NH$_3$ with ND$_3$ should bring the transition
pressure down by a factor of 15, and replacing hydrogen with a gas of some
other nonpolar molecule might be advantageous.  Thus a direct experimental
test of whether dipole-dipole interactions are or are not essential for
understanding the vanishing of the tunneling frequency seems feasible.

Our very simple decoherence model corresponds to setting $\gm=\zt$ and $\dl=0$
in Ben-Reuven's theory as it applies to a two-level system.  In fact, he
achieved a good fit to the experimental data with $\dl=0$, but with $\gm$
larger than $\zt$ by a factor of around 1.3, see p.~21 of \cite{BRvn66}.  To
have $\gm$ larger than $\zt$ in our model would require our adding another
source of decoherence. The phase transition present in our model is also
clearly present in Ben-Reuven's work; see the discussion of the spectrum of
the perturbed Liouville matrix in Sec.~4C of \cite{BRvn66}, where the
eigenvalues change character when $\zt$ passes through the value $\om_0+\dl$;
this is the counterpart of our $\gm=\om$. Hence it seems that the vanishing of
the inversion frequency in ammonia with increasing pressure is an instance of
the sort of phase transition that occurs in our model. A more detailed
comparison, which we have not attempted, would require our including an
additional mechanism for decoherence to make $\gm$ larger than $\zt$, and
dealing with complications caused by the presence in ammonia of a number of
different rotational states.  Nonetheless, we think our considerations provide
some insight into the sense in which the ammonia molecule can be said to be
``pyramidal'' in the gas at high pressure and lack this feature at low
pressures. Namely, when collisions are sufficiently frequent there is a
consistent family of histories, the chiral or $x$ family discussed in
Sec.~\ref{sbct4.5}, in which in quantum mechanical terms the molecule
spends a time much longer than the tunneling time in a pyramidal shape (or a
``dressed'' state close to it) which is ``chiral'' in the sense that electric
dipole moment has a definite orientation relative to the angular momentum,
with occasional random hops between the two pyramidal possibilities.  As the
pressure decreases towards the transition pressure the pyramidal picture
begins to break down: the hops become more frequent between dressed states,
which are starting to lose their pyramidal character. At still lower pressures
it is better to think of the molecule as continuously tunneling back and
forth, rather than possessing a fixed pyramidal form, with a period that
diverges as the pressure rises to its value at the transition.

\xb
\section{\  Conclusion \label{sct7}}
\xa

We have shown how the decoherence of a two-state tunneling molecule, a chiral
molecule or ammonia, in the presence of a buffer gas can be described in terms
of a succession of quantum states of the molecule itself that form a
consistent family of histories, on a sufficiently coarse time scale so that
intervals are always longer than a correlation time.  Our model is described
by just two parameters, a tunneling rate $\omega$ and a decoherence rate
$\gamma$, and its essential properties depend upon the ratio $\gm/\om$.  In
addition we have studied the flow of information to the environment, along with
its retention by the molecule itself, during the process of decoherence.

We found a large variety of consistent families, some of which are stationary
(in the sense of Markov processes) and some of which are not. In the regime
$\gm/\om\gg 1$ of strong decoherence there is a stationary family (actually
two closely related families) in which the molecule spends a relatively long
time in one of its chiral states before flipping to the other chirality, and
eventually flipping back again, in a stationary Markov (``telegraph'')
process, with a transition rate which is approximately $\om^2/4\gm$ for $\gm
\gg \om$, and hence quite slow compared to the tunneling rate $\om$.  Thus this
``chiral'' family explains the persistence of chirality for a long period
of time when there is strong decoherence.  However, as $\gm/\om$ decreases,
the transitions between chiral states become more frequent and the states
themselves (the ``dressed $x$'' states of Sec.~\ref{sbct4.2}) lose their
chiral character, until finally this family disappears entirely at a
phase transition $\gm/\om=1$.

We have found two other stationary consistent families for $\gm/\om > 1$.  One
of them is the ``parity'' family in which the molecule is at each of the times
considered in one of the two states of definite parity (the energy eigenstates
of the isolated molecule), but with a transition rate of $\gm$ between them,
thus a rapid flipping compared to transitions between chiral states when the
decoherence is large.  This family, present at all values of $\gm/\om$, is
incompatible, in the quantum mechanical sense, with the chiral family: while
both provide valid quantum descriptions, they cannot be employed
simultaneously; see the discussion in Sec.~\ref{sbct4.5}.  The other
stationary family for $\gm/\om > 1$ (again there are actually two families) is
the ``dressed $y$'' family of Sec.~\ref{sbct4.2}. It involves a relatively rapid
flipping between two orthogonal quantum states for which we do not have a
simple physical interpretation.  Like the chiral family this one only exists
for $\gm > \om$.

For $\gm < \om$ there is a nonstationary ``tunneling'' family in which the
molecule oscillates back and forth between the two chiral states at a rate
that goes to zero as $\gm/\om$ increases to 1.  On top of this relatively
smooth oscillation there are random changes in phase which constitute a
nonstationary Markov process, with a rate that increases with $\gm$.  This
tunneling family disappears at the phase transition $\gm=\om$.  The only truly
stationary family in the regime $\gm < \om$ is the parity family.  In
addition, both for $\gm < \om$ and for $\gm > \om$ there are a variety of
nonstationary consistent families in which the orthogonal basis used to describe
the quantum system tends with time towards one of the stationary families or,
for $\gm < \om$, the tunneling family.

It seems likely that most chiral molecules under most conditions will be in a
regime of strong decoherence $\gm \gg \om$, see the remarks about D$_2$S$_2$
in Sec.~\ref{sbct6.1}. Whereas it can only be thought of as ``chiral'' when in
an appropriate rotational state, ammonia, including its deuterated form
ND$_3$, is an example of a tunneling molecule in which the
transition at $\gm/\om=1$ can be readily observed in the laboratory.  Indeed,
it appears that it has already been observed; see the discussion in
Sec.~\ref{sbct6.2}.  One wonders if additional experiments, perhaps using
technques other than, or in addition to, microwave absorption might be helpful
in elucidating its behavior near the phase transtion.

In addition to consistent quantum families of histories we have studied,
within the scope of our simple model, the flow of information from a tunneling
molecule to its environment, along with the loss of information in the
molecule itself.  In Sec.~\ref{sct5} we used a perspective in which at a later
time the information about the quantum state of the molecule at an earlier
time is thought of as a quantum channel, while similar information present at
this later time in the environment constitutes a complementary channel.  What
happens in both cases depends strongly on the type of information
considered. Given that our model of decoherence, Sec.~\ref{sbct2.2}, is based
on the flow of chiral ($X$) information---is the molecule left or right
handed?---to the environment, we were not surprised to find this exhibited in
our quantitative measures, together with a  rapid decrease of
``complementary'' types of information, corresponding to bases mutually
unbiased with respect to $X$, retained within the molecule itself.  Indeed,
there is a direct quantitative relationship for short times if one uses a
Holevo type of information measure, and an exact equality in the instantaneous
rates, \eqref{eqn47}, if in the Holevo measure von Neumann entropy is replaced
with  quadratic entropy in order to render the rates finite.

There are a number of ways in which lines of investigation initiated in this
paper could be further extended. Parity-violation effects could be modeled as a small energy splitting between chiral states \cite{QuackReview2002}. Also, our model contains only one mechanism for decoherence, Sec.~\ref{sct2}; adding a second would
allow a serious comparison with Ben-Reuven's formula for ammonia as discussed
in Sec.~\ref{sbct6.2}.  Obtaining the correct physical interpretation might
prove difficult given the complexity of the rotational states, even for
ammonia present as a dilute component in a buffer gas.  Fluorescence from a
two-level atom, where decoherence arises from spontaneous decay, could
be a simpler system for studying the dynamical phase transition. This transition has been studied
in terms of correlations among scattered photons in \cite{GrLs10}, and it
would be of interest to supplement this with a description of how the atom
itself behaves as a function of time.  Indeed, even the decay of an isolated
atom initially in an excited state has not, so far as we know, been examined
using consistent families, and studying them might yield valuable physical
insights.

\section*{Acknowledgments}
The research described here was supported by the Office of Naval Research. V. Gheorghiu acknowledges additional support from the Natural Sciences and Engineering Research Council of Canada (NSERC) and  from a Pacific Institute for Mathematical Sciences (PIMS) postdoctoral fellowship.

\appendix

\xb
\section{Explicit expressions for $\Tmat$ and its eigenvectors\label{apdxA}}
\xa

The general expression for $\Tmat(t)=e^{t\Smat}$ in \eqref{eqn7} is
\begin{equation}
  \Tmat(t)=\begin{pmatrix}
 1 & 0              & 0               & 0 \\
 0 & e^{-\gm t}a(t) & -e^{-\gm t}b(t) & 0 \\
 0 & e^{-\gm t}b(t) & -e^{-\gm t}c(t) & 0 \\
 0 & 0              & 0               & e^{-2\gm t} \\
            \end{pmatrix}
\label{eqn50}
\end{equation}
where, for $\gm > \omt$ 
\begin{align}
a(t) &= \cosh \xi t +(\gm/\xi)\sinh \xi t
\notag\\
b(t) &= (\omt/\xi)\sinh \xi t
\notag\\
c(t) &= \cosh \xi t -(\gm/\xi)\sinh \xi t
\notag\\
\xi &:= \sqrt{\gm^2-{\omt}^2}, 
\label{eqn51}
\end{align}
whereas for $\gm < \omt$
\begin{align}
a(t) &= \cos \eta t +(\gm/\eta)\sin \eta t
\notag\\
b(t) &= (\omt/\eta)\sin \eta t
\notag\\
c(t) &= \cos \eta t -(\gm/\eta)\sin \eta t
\notag\\
\eta &:= \sqrt{{\omt}^2-\gm^2}. 
\label{eqn52}
\end{align}
For $\gm = \omt$ one has
\begin{equation}
  a(t) = 1+\gm t,\quad b(t) = \gm t,\quad c(t) = 1-\gm t,
\label{eqn53}
\end{equation}
where of course $\gm$ could be replaced by $\omt$. 

For the eigenvalues $\lm_1$ and $\lm_4$ in \eqref{eqn14} the left and right
eigenvectors of $\Tmat$ are, trivially, $(1,0,0,0)$ and $(0,0,0,1)$,
respectively. For $\lm_2$ and $\lm_3$, the unnormalized left $\vect{v}$ and
(transposed) right $\vect{w}$ eigenvectors are:
\begin{alignat}{2}
\vect{v}_2 &= (0,\xi - \gm,\omt,0),\quad &
\vect{w}_2 &= (0,\gm -\xi,\omt,0), 
\notag\\
\vect{v}_3 &= (0,-\gm-\xi,\omt,0), \quad &
\vect{w}_3 &= (0,\gm +\xi,\omt,0), 
\label{eqn54}
\end{alignat}
where for $\gm < \omt$ replace $\xi$ with $i\eta$.

For short times $t \ll1$ one has 
 \begin{equation}\label{eqn55}
\Tmat(t)=\begin{pmatrix}
1 & 0 & 0 & 0\\
0 & 1 & -\omt t & 0\\
0 & \omt  t & 1-2\gamma t & 0\\
0 & 0 & 0 & 1-2\gamma t
\end{pmatrix}+\Omat( t^2).
\end{equation}
for the direct channel and
\begin{align}\label{eqn56}
\Tmat^c(t)=&
\begin{pmatrix}
1 & 0 & 0 & 0\\
0 & 2\sqrt{\gamma}\sqrt{ t} -\gamma^{3/2} t^{3/2} & \omt\sqrt{\gamma} t^{3/2} & 0\\
0 & 0 & 0 & 0\\
1-2\gamma t & 0 & 0 & 0
\end{pmatrix}\notag\\
&+\Omat( t^2).
\end{align}
for the complementary channel. During this short time interval, one can represent the direct channel $\TC$ using using only 2 Kraus operators, instead of 4 required by the most general qubit channel. The physical intuition behind this is that during this time interval the system of interest interacts only with a qubit environment, being effectively ``decoupled'' from the other environmental qubit (we remind the reader that the most general qubit evolution requires an interaction with an environment that is represented by at least 2 qubits, see e.g. \cite{NielsenChuang:QuantumComputation}).

\xb
\section{Differential equations for consistent families\label{apdxBB}}
\xa

As discussed in Sec.~\ref{sbct4.1}, the forwards and backwards consistency
conditions specify how a diameter of the Bloch sphere rotates.  To determine
this for the forwards condition, consider the density operator which at the
initial time is at one end of the diameter, and write it in the form
\begin{equation}
  \rho = {\textstyle\frac{1}{2} }(I + \vect{r}\cdot\vect{\sg}),\quad
\vect{r} = r\vect{n},
\label{eqn57}
\end{equation}
using the notation of \eqref{eqn43} and \eqref{eqn44}. Because $\TC $ is
unital the master equation \eqref{eqn12} for $\bld{\rho}$ is equivalent to
\begin{equation}
  d\vect{r}/dt = \Sbmat\cdot\vect{r},
\label{eqn58}
\end{equation}
or
\begin{equation}
\Bigl(\frac{dr}{dt}\Bigr)\vect{n} + \Bigl[ r\frac{d\vect{n}}{dt}\Bigr]
=\Bigl(\vect{n}\cdot\Sbmat\cdot\vect{n}\Bigr)\vect{n} + 
\Bigl[\Sbmat\cdot\vect{r} - 
\left(\vect{n}\cdot\Sbmat\cdot\vect{n}\right)\vect{n}\Bigr],
\label{eqn59}
\end{equation}
where $\Sbmat$ is the lower right $3\times 3$ block of $\Smat$ in 
\eqref{eqn7}.  Set  $r=1$, thus $\vect{r} =\vect{n}$, and take the dot product
of both sides of \eqref{eqn59} with $\vect{n}$, noting that
$\vect{n}$ and $d\vect{n}/dt$ are necessarily orthogonal to each other, to
obtain:
\begin{align}
  dr/dt &= \vect{n}\cdot\Sbmat\cdot\vect{n},
\label{eqn60}
\\
  d\vect{n}/dt &= \Sbmat\cdot\vect{n} 
- (\vect{n}\cdot\Sbmat\cdot\vect{n})\vect{n}.
\label{eqn61}
\end{align}
Note that $\vect{n}\cdot\Sbmat\cdot\vect{n}$ depends only on the symmetrical
part of $\Sbmat$, which is to say the dissipative term, proportional to $\gm$,
in the master equation \eqref{eqn13}.
The differential equations in \eqref{eqn30} are equivalent to \eqref{eqn61}
when $\vect{n}$ is written in polar coordinates.  To obtain the differential
equations for the backwards
consistency condition, replace $\Sbmat$ in \eqref{eqn61} with $\Sbmat\ad$,
corresponding to the adjoint superoperator $\TC\ad$, and $d/dt$ with $-d/dt$.
The resulting differential equations are equivalent to \eqref{eqn30} with 
$\gm$ replaced with $-\gm$, thus  \eqref{eqn33}.

The consistency conditions are related to the motion in the Bloch sphere of
the diameter that corresponds to the (instaneous) orthonormal basis. However,
the instantaneous hopping rate $\kp$ for the Markov process can be calculated
using the Born rule for a very short time interval during which one can assume
that the diameter remains fixed, as its motion (the change in basis) only
contributes to higher order. When $r=1$, $\kp$ as defined in \eqref{eqn32} is
equal to $(-1/2)dr/dt$. Thus,  using \eqref{eqn60},
\begin{equation}
  \kp =(-1/2) (\vect{n}\cdot\Sbmat\cdot\vect{n})
=\gm(1-n_x^2),
\label{eqn62}
\end{equation}
which, transformed to polar coordinates, is \eqref{eqn31}.

\begin{widetext}
\xb
\section{Time-integrated solutions for consistent families\label{apdxB}}
\xa

It is helpful to define $\mu (t):=\tan \phi(t)$ and $\nu(t):=\tan \th(t)$. Then 
\eqref{eqn30} and \eqref{eqn33} can be integrated to give the following 
explicit solutions for the non-stationary consistent families:
\begin{align}
\mu(t)&=\mu(0)+(\omt \mp 2\gm \mu(0)+
\omt \mu(0)^2)\frac{\sinh \xi t}{\xi \cosh \xi t +(\pm \gm -\omt \mu(0))\sinh \xi t}\notag\\
\nu(t)&=\nu(0)e^{\pm \gm t}\sqrt{1\pm (\gm/\xi)
\frac{1-\mu(0)^2}{1+\mu(0)^2} \sinh 2 \xi t +(2\gm/\xi^2)
\Bigl(\gm \mp \frac{2\omt  \mu(0)}{1+\mu(0)^2} \Bigr) \sinh^2 \xi t
}
\label{eqn63}
\end{align}
where the top (or bottom) symbol in $\pm$ or $\mp$ is the solution to
\eqref{eqn30} (or \eqref{eqn33}) for the family satisfying the forward
(or backward) condition. In the case that $\omt >\gm$, one can replace every occurrence of $\xi=\sqrt{\gm^2-\omt^2}$ in \eqref{eqn63} with $\eta=\sqrt{\omt^2-\gm^2}$, provided that $\sinh$ and $\cosh$ are replaced by $\sin$ and $\cos$.

\xb
\section{Equality of mutual information and $\chi$ measure when
  forwards consistency conditions satisfied \label{apdxC}}
\xa

The key observation is that, when the forward condition is satisfied, $\TC_{m,1}(P^j_1)=\sum_k q_{kj} P^k_m$ for each $j$ and hence the ensemble of density operators at the channel output commute with each other, so quantum (von Neumann) entropies of these density operators become classical (Shannon) entropies in the basis that diagonalizes these density operators. Denoting $r^j_1:= \Tr(P_1^j)$ and $r^k_m:= \Tr(P_m^k)$, we have 
\begin{align}
\hat\chi(P_1,\TC_{m,1})=& S(\sum_j p_j \TC_{m,1}(P^j_1)/r^j_1)-\sum_j p_j S(\TC_{m,1}(P^j_1)/r^j_1)\notag\\
=&S(\sum_{j,k} p_j q_{kj} P^k_m /r^j_1)-\sum_j p_j S(\sum_k q_{kj} P^k_m/r^j_1) \notag\\
=&H(\{ \sum_j p_jq_{kj} r^k_m/ r^j_1 \}_k)+ \sum_{j,k} (p_j q_{kj} r^k_m /r^j_1) S(P^k_m /r^k_m) \notag\\
&-\sum_j p_j H(\{ q_{kj} r^k_m /r^j_1 \}_k ) - \sum_{j,k} (p_j q_{kj} r^k_m /r^j_1) S(P^k_m /r^k_m) \notag\\
=&H(\{ \sum_j p_jq_{kj} r^k_m/ r^j_1 \}_k)-\sum_j p_j H(\{ q_{kj} r^k_m /r^j_1 \}_k ) \notag\\
=&H(P_m)-H(P_m |P_1 )=H(P_1\colo P_m), 
\label{eqn64}
\end{align}
where the $k$ subscript in $\{\cdot \}_k$ indicates that the set is generated by allowing $k$ to vary. In this derivation, we used a property of the von Neumann entropy, for orthogonal density operators, given on page~513 of \cite{NielsenChuang:QuantumComputation}. 
\end{widetext}
\xb


\end{document}